\documentclass[runningheads]{llncs}
\usepackage[T1]{fontenc}
\usepackage{amsfonts}
\usepackage{esvect}
\usepackage{multirow}
\usepackage{url}

\usepackage{hyperref}

\hypersetup{
  hidelinks,
  colorlinks=true,
  allcolors=blue,
  pdfstartview=Fit,
  breaklinks=true
}

\newcommand{\subfour}[1]{\vspace*{3mm}{\noindent\bf #1}} 

\usepackage{booktabs}
\usepackage{graphicx}
\usepackage{xcolor}
\usepackage{soul}
\usepackage[numbers,sort&compress]{natbib}
\let\oldmaketitle\maketitle
\renewcommand{\maketitle}{\oldmaketitle\setcounter{footnote}{0}}

\begin{document}
\title{Self-Supervised Contrastive BERT Fine-tuning for Fusion-based Reviewed-Item Retrieval}
\titlerunning{Self-Supervised Contrastive BERT Fine-tuning for Fusion-based RIR} 

%
\author{Mohammad Mahdi Abdollah Pour\inst{1}$^*$ \and
Parsa Farinneya\inst{1}$^*$ \and
Armin Toroghi\inst{1}
\and
Anton Korikov\inst{1}
\and
Ali Pesaranghader\inst{2}
\and
Touqir Sajed\inst{2}
\and
Manasa Bharadwaj\inst{2}
\and
Borislav Mavrin\inst{2}
\and
Scott Sanner\inst{1}
}
\def\thefootnote{*}\footnotetext{Equal Contribution}\def\thefootnote{\arabic{footnote}}

\authorrunning{M. Abdollah Pour et al.}

\institute{University of Toronto \and LG Electronics, Toronto AI Lab
\email{\{m.abdollahpour,parsa.farinneya,armin.toroghi,anton.korikov\}@mail.utoronto.ca }  
\email{\{ali.pesaranghader,touqir.sajed,manasa.bharadwaj,borislav.mavrin\}@lge.com}
\email{ssanner@mie.utoronto.ca}}
 \maketitle              
%
\begin{abstract}
As natural language interfaces enable users to express increasingly complex natural language queries, there is a parallel explosion of user review content that can allow users to better find items such as restaurants, books, or movies that match these expressive queries.
While Neural Information Retrieval (IR) methods have provided state-of-the-art results for matching queries to documents, they have not been extended to the task of Reviewed-Item Retrieval (RIR), where query-review scores must be aggregated (or fused) into item-level scores for ranking.
In the absence of labeled RIR datasets, we extend Neural IR methodology to RIR by leveraging self-supervised methods for contrastive learning of BERT embeddings for both queries and reviews.
Specifically, contrastive learning requires a choice of positive and negative samples, where the unique two-level structure of our item-review data combined with meta-data affords us a rich structure for the selection of these samples. For contrastive learning in a Late Fusion scenario (where we aggregate query-review scores into item-level scores), 
we investigate the use of positive review samples from the same item and/or with the same rating, selection of hard positive samples by choosing the least similar reviews from the same anchor item, and selection of hard negative samples by choosing the most similar reviews from different items.
We also explore anchor sub-sampling and augmenting with meta-data.  
For a more end-to-end Early Fusion approach, we introduce contrastive item embedding learning to fuse reviews into single item embeddings. 
Experimental results show that Late Fusion contrastive learning for Neural RIR  
outperforms all other contrastive IR configurations, Neural IR, and sparse retrieval baselines, thus demonstrating the power of exploiting the two-level structure in Neural RIR approaches as well as the importance of preserving the nuance of individual review content via Late Fusion methods.

\end{abstract}
\section{Introduction}
\begin{figure}[t]
    \centering
    \includegraphics[scale=0.3]{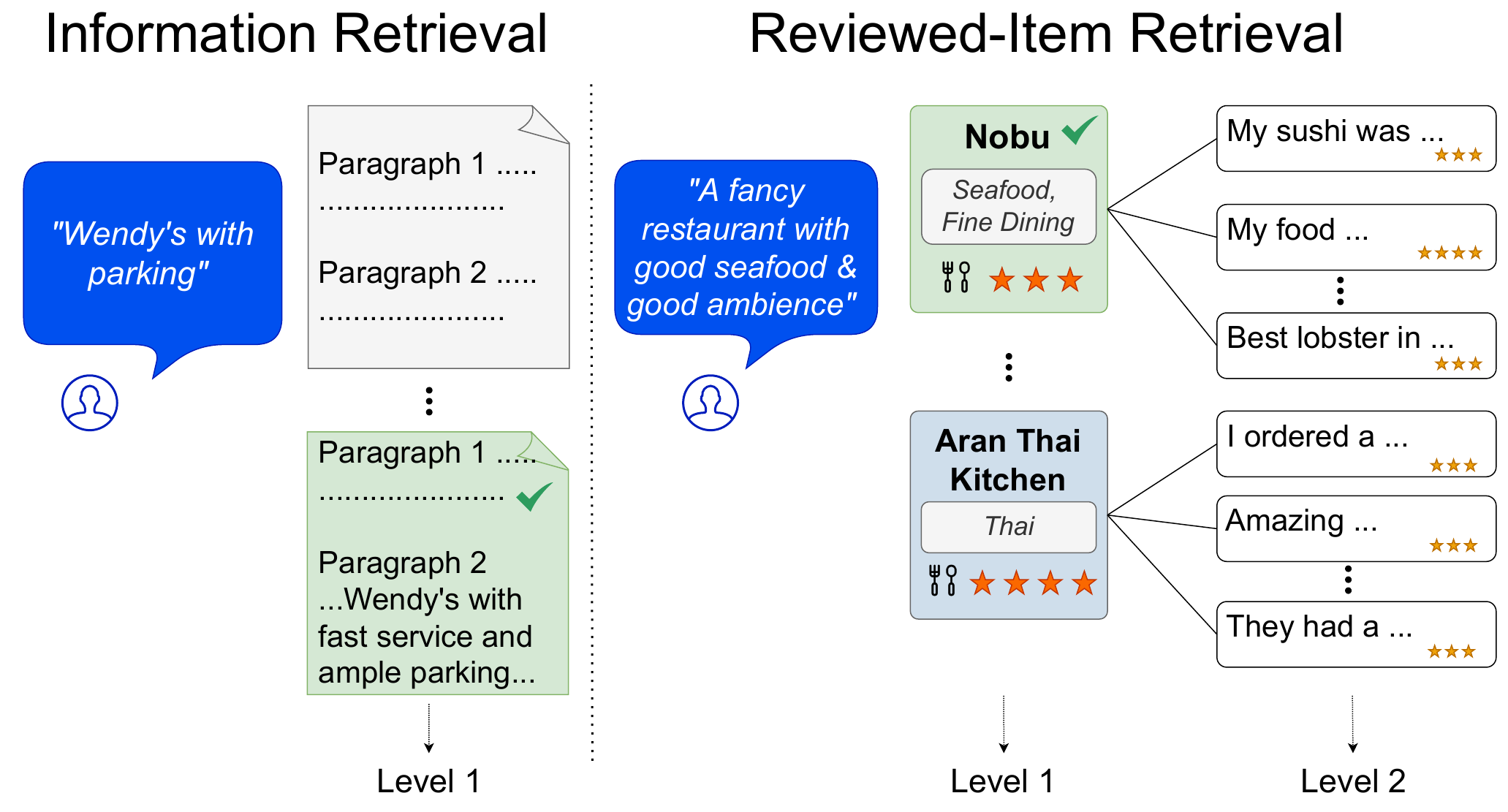}
    \vspace{-2mm}
    \caption{Structural difference between IR (left) and RIR (right). In RIR, items have reviews covering different aspects. In contrast, documents in the IR task do not have this \textbf{two-level item-review structure}. 
    }
    \vspace{-4mm}
    \label{pmvsir}
\end{figure}

The rise of expressive natural language interfaces coupled with the prevalence of user-generated review content provide novel opportunities for query-based retrieval of reviewed items.  While Neural Information Retrieval (IR) methods have provided state-of-the-art results for query-based document retrieval~\citep{lin2021pretrained}, these methods do not directly extend to 
review-based data that provides a unique two-level structure in which an item has several reviews along with ratings and other meta-data. Due to differences between standard IR document retrieval~\citep{schutze2008introduction} and the task of retrieving reviewed-items indirectly through their reviews, we coin the term Reviewed-Item Retrieval (RIR) for this task. 
Figure\ \ref{pmvsir} illustrates the structural difference between IR and RIR. In RIR, each item includes a set of reviews and each review expresses different perspectives about that item. 

Unlike standard Neural IR, which produces query-document scores for document ranking, RIR requires query-item scores to be obtained indirectly using review text. This can be done both via Late Fusion (LF) methods that simply aggregate query-review scores into item scores or Early Fusion (EF) methods that build an item representation for direct query-item scoring.  Given the absence of labeled data for RIR, we explore self-supervised contrastive learning methods for Neural RIR that exploit the two-level structure of the data in both the LF and EF frameworks with the following contributions:
\begin{enumerate}

    \item For LF, we propose different positive and negative sampling methods that exploit the two-level structure of RIR data as well as data augmentation methods for contrastive fine-tuning of a  BERT~\cite{devlin2018bert} Language Model (LM).

\item For EF, we propose end-to-end contrastive learning of item embeddings to fuse reviews of each item into a single embedding vector.

\item 
We experimentally show that LF-based Neural RIR methods outperform EF and other
contrastive IR methods, Neural IR, and sparse retrieval
baselines.
\end{enumerate}
Overall, these results demonstrate the power of exploiting the two-level structure in Neural RIR approaches. By showing the superiority of LF over EF, we also demonstrate the benefit of aggregating scores for individual reviews \textit{after} similarity with a query is computed, thus preserving the nuances of each review during scoring. Ultimately, this work develops a foundation for the application and extension of Neural IR techniques to RIR.

\section{Background}

\subsection{IR}
Given a set of documents $\mathcal{D}$ and a query $q \in \mathcal{Q}$, an IR task $\mathcal{IR}\langle \mathcal{D},q \rangle$ is to assign a similarity score $S_{q,d} \in \mathbb{R}$ between the query and each document $d \in \mathcal{D}$ and return a list of top-scoring documents. 

Before the advent of Neural IR methods, most methods depended on sparse models such as TF-IDF \citep{salton1975vector} and its variants such as BM25 \cite{robertson2009probabilistic},  which heavily relied on exact term matches and measures of term informativeness. However, the need for exact term matches,
the availability of large datasets, and increases in computational power led to a shift from traditional models to deep neural networks for document ranking \cite{10.1145/2505515.2505665,10.1145/2600428.2609622}.
Recently, \citet{nogueira2019passage}
have initiated a line of research on Neural IR by fine-tuning BERT~\cite{devlin2018bert} for ranking candidate documents with respect to queries.

Recent 
works have substantially extended BERT-based Neural IR methods. CoCondenser \cite{gao2022unsupervised} is a resource-efficient model with excellent performance on the MS-MARCO benchmark \citep{nguyen2016ms}. This model is based on Condenser \citep{gao2021condenser}, which alters the BERT architecture to emphasize more attention to the classification embedding (i.e., the so-called ``[CLS]'' output of BERT).
Contriever \citep{izacard2021towards} is a  state-of-the-art self-supervised contrastive Neural IR method that does not rely on query-document annotations to train. 

Although our work on Neural RIR is influenced by Neural IR, the structure of IR and RIR differ as illustrated by Figure \ref{pmvsir}. This requires methods specifically for working with the two-level structure of data in both training and inference in the RIR task. To the best of our knowledge, these methods have not been explored in the literature.

\subsection{Fusion} \label{fusion}
Information retrieval from two-level data structures has previously been studied by Zhang and Balog \cite{zhang2017design}, though they did not study \textit{neural} techniques, which are the focus of our work. Specifically, Zhang and Balog define the \textit{object retrieval} problem, where (high-level) objects are described by multiple (low-level) documents and the task is to retrieve objects given a query. This task requires \textit{fusing} information from the document level to the object level, which can be done before query scoring, called Early Fusion, or after query scoring, called Late Fusion. Our contributions include extending Early and Late Fusion methods to self-supervised contrastive Neural IR.

Formally, let $i \in \mathcal{I}$ be an object described by a set of documents $\mathcal{D}_i \subset \mathcal{D}$ and let $r_{i,k}$ denote the $k$'th document describing object $i$. Given a query $q \in \mathcal{Q}$, fusion is used to aggregate document information to the object level and obtain a query-object similarity score $S_{q,i} \in \mathbb{R}$.

\subsubsection{Late Fusion}
\label{late_fusion_bg}

\label{late_fusion}
In Late Fusion, similarity scores are computed between documents and a query and then aggregated into a query-object score. 
Given an embedding space $\mathbb{R}^m$, let $g: \mathcal{D} \cup \mathcal{Q} \rightarrow \mathbb{R}^m$ map $r_{i,k}$ and $q$ to their embeddings $g(r_{i,k}) = \vec{r}_{i,k}$ and $g(q) = \vec{q}$, respectively.  Given a similarity function $f(\cdot,\cdot):\mathbb{R}^m \times \mathbb{R}^m \rightarrow \mathbb{R}$, a query-document score $S_{q,r_{i,k}}$ for document $r_{i,k}$ is computed as
\begin{equation}
\label{query_review}
    S_{q,r_{i,k}} = f(\vec{q}, \vec{r_{i,k}}) 
\end{equation}
To aggregate such scores into a similarity score for object $i$, the top-$K$ query-document scores for that object are averaged: 

\begin{equation}
\label{review_fusion}
  S_{q,i} = \frac{1}{K}\sum_{j=1}^{K} S_{q,r_{i,j}}
\end{equation}
where $S_{q,r_{i,j}}$ is the $j$'th top query-document score for object $i$. The Late Fusion process is illustrated on the left side of Figure \ref{latefusion}.

Previously, Bursztyn \textit{et al.} \cite{bursztyn-etal-2021-doesnt} introduced a method (one of our baselines) which can be interpreted as neural late fusion with $K = 1$, in which query and document embeddings are obtained from an LM fine-tuned using conventional Neural IR techniques. 
In contrast, we develop neural late fusion methods where $K \neq 1$ and introduce novel contrastive LM fine-tuning techniques for neural fusion.


\subsubsection{Early Fusion}
\label{early_fusion}
\begin{figure*}[h!]
\centering
\includegraphics[scale=0.21]{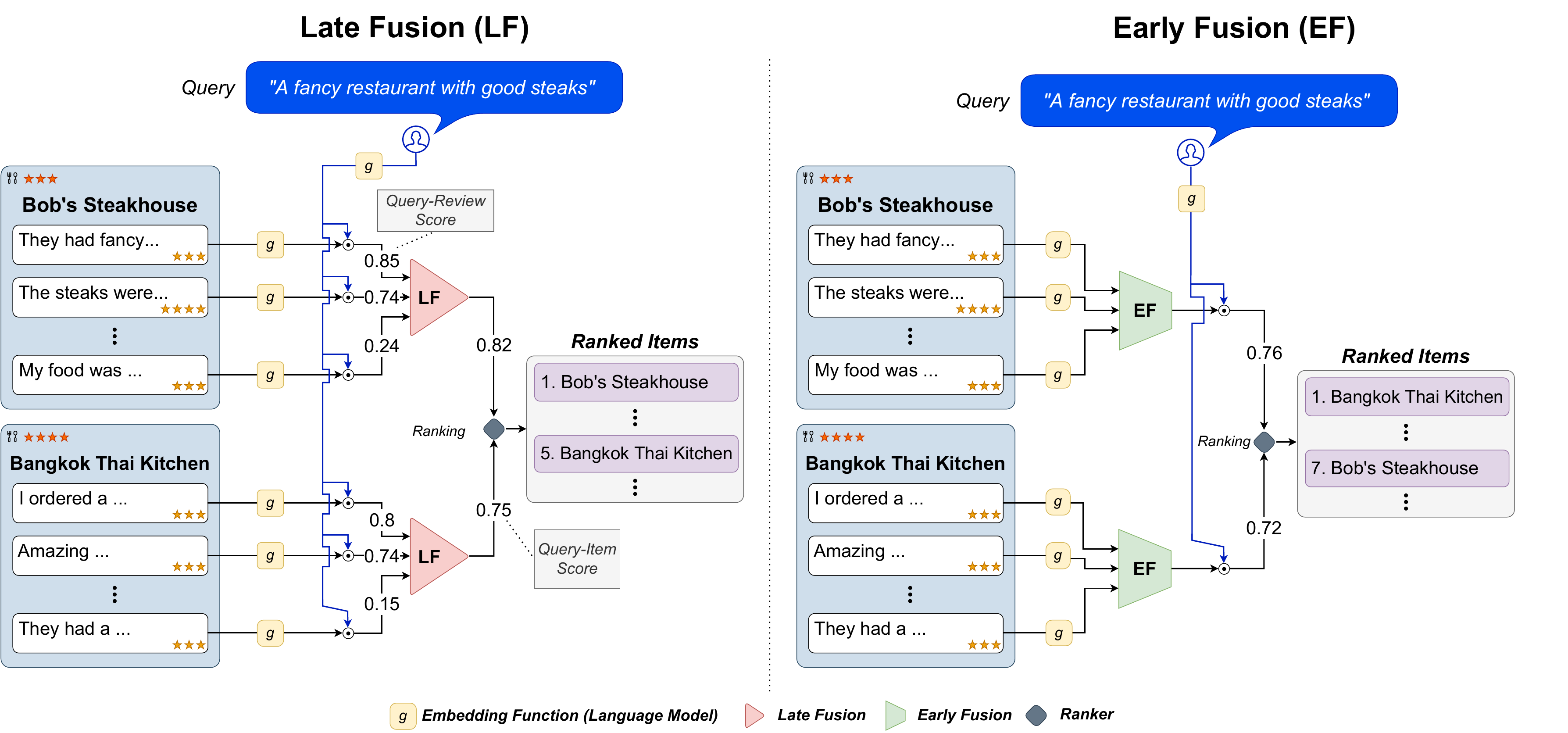}
\caption{(Left) Late Fusion demonstrates how embedding of queries and reviews are used to get a similarity score with Equation (\ref{query_review}) and how 
items are ranked according to the query-item score. (Right) Early Fusion demonstrates how reviews are fused together to build a vector representation for comparing the item with the query. 
}
\vspace{-4mm}
\label{latefusion}
\end{figure*}

In Early Fusion, document information is aggregated to the object level before query scoring takes place. Various aggregation approaches are possible and we discuss the details of our proposed methods in Section \ref{early}; however, the purpose of aggregation is to produce an object embedding $\vec{i} \in \mathbb{R}^m$. Query-item similarity is then computed directly as 
\begin{equation}
\label{ef_formula}
    S_{q,i} = f(\vec{q},\vec{i}) 
\end{equation}
The Early Fusion process is illustrated on the right side of Figure \ref{latefusion}.

Interestingly, related work \cite{yang2016hierarchical,shing2020prioritization} which uses hierarchical neural networks to classify hierarchical textual structures can be interpreted as Supervised Early Fusion, since the hierarchical networks learn to aggregate low-level text information into high-level representations. In contrast to these works, we study retrieval, use self-supervised contrastive learning, and explore both Early and Late Fusion.

\subsection{Contrastive Representation Learning} \label{sec:revCL}
To generate embeddings, our methods rely on contrastive representation learning. Given data samples $x \in \mathcal{X}$
 where each $x$ is associated with a label,
the goal of contrastive learning is to train an embedding function $g:\mathcal{X} \rightarrow \mathbb{R}^m$ which maximizes the similarity between two embeddings $f(g(x^a),g(x^b)) = f(\vec{x}^a, \vec{x}^b)$ if $x^a$ and $x^b$ are associated with the same label, and minimizes this similarity if $x^a$ and $x^b$ are associated with different labels. During training, these similarities are evaluated between some anchor $x^A \in \mathcal{X}$, positive samples $x^+ \in \mathcal{X}$ associated with the same label as $x^A$, and negative samples $x^- \in \mathcal{X}$ associated with different labels than $x^A$. Specifically, given a tuple of $N+1$ inputs $(x^A,x^+,x^-_1,...,x^-_{N-1})$, a similarity function $f$, and an embedding function $g$, the N-pair loss \cite{sohn2016improved} is defined as:
\begin{equation}
\label{npair}
 L^g_{\mathrm{con}}(x^A,x^+,\{x_i^-\}_{i=1}^{N-1})   = - \mathrm{log} \frac{e^{f(g(x^{A}),g(x^+))}}{e^{f(g(x^{A}),g(x^+))} + \sum_{i=1}^{N-1} e^{f(g(x^{A}),g(x^-_i))}} 
\end{equation}
This equation is equivalent to the softmax loss for multi-class classification of $N$ classes. Letting $s = (x^A, x^+, \{x_i^-\}_{i=1}^{N-1} )$ denote a sample tuple and $\mathcal{S}$ be the set of samples used for training, the objective of contrastive learning is:
\begin{equation}
    \min_g L^g(\mathcal{S}) = \min_g \sum_{s \in \mathcal{S}}  L^g_{\mathrm{con}}(s) \label{eq:totalLoss}
\end{equation}


\subsubsection{Sampling Methods for Neural IR
}
\label{IRcon}

For contrastive learning to be used for Neural IR, a set of samples $\mathcal{S}$ must be created from the documents. We focus on two sampling methods for this task as baselines. The \textit{Inverse Cloze Task} (ICT) \citep{izacard2021towards} uses two mutually exclusive spans of a document as a positive pair $(x^A,x^+)$ and spans from other documents as negative samples. \textit{Independent Cropping} (IC) \citep{izacard2021towards,gao2022unsupervised} takes two random spans of a document as a positive pair and spans from other documents as negative samples. Another sampling approach used in ANCE \cite{xiong2020approximate} relies on query similarity to dynamically identify additional negative samples, but we do not compare to this approach since we do not have access to queries in our self-supervised learning setting.


\subsubsection{In-Batch Negative Sampling}
\label{IB}
Equation (\ref{eq:totalLoss}) is often optimized using mini-batch gradient descent with in-batch negative sampling \cite{yih2011learning,henderson2017efficient,gillick2019learning,karpukhin2020dense}. Each mini-batch contains $N$ anchors and $N$ corresponding positive samples. The $j$'th sample tuple $s_j$ consists of an anchor $x_j^A$, a positive sample $x_j^+$, and, to improve computational efficiency, the positive samples of other tuples $s_{j',j' \neq j}$ are used as negative samples for tuple $s_j$. That is, the set of negative samples for $s_j$ is  $\{x^+_{j'}\}_{j' = 1, j' \neq j}^N$.


\section{Proposed Fusion-based Methods for RIR}

We now define the Reviewed-Item Retrieval problem as a specific and highly applicable case of object retrieval (Section \ref{fusion}). We then demonstrate how the two-level structure of reviewed-item data can be exploited by our novel contrastive fine-tuning methods for late and early fusion. In the Reviewed-Item Retrieval problem $\mathcal{RIR} \langle \mathcal{I},\mathcal{D},q \rangle$, we are given a set of $n$ items $\mathcal{I}$ where each item $i \in \mathcal{I}$ is described by a set of reviews $\mathcal{D}_i \subset \mathcal{D}$, and where the $k$'th review of item $i$ is denoted by $r_{i,k}$. A review $r_{i,k}$ can only describe one item $i$, and this requirement makes RIR a special case of object retrieval since, in object-retrieval, a document can be associated with more than one object. Given a query $q \in \mathcal{Q}$, the goal is to rank the items based on the $S_{q,i}$ score for each item-query pair $(q,i)$.

\subsection{CLFR: Contrastive Learning for Late Fusion RIR}
\label{CLFR}
\begin{figure*}[t!]
\centering
\includegraphics[scale=0.45]{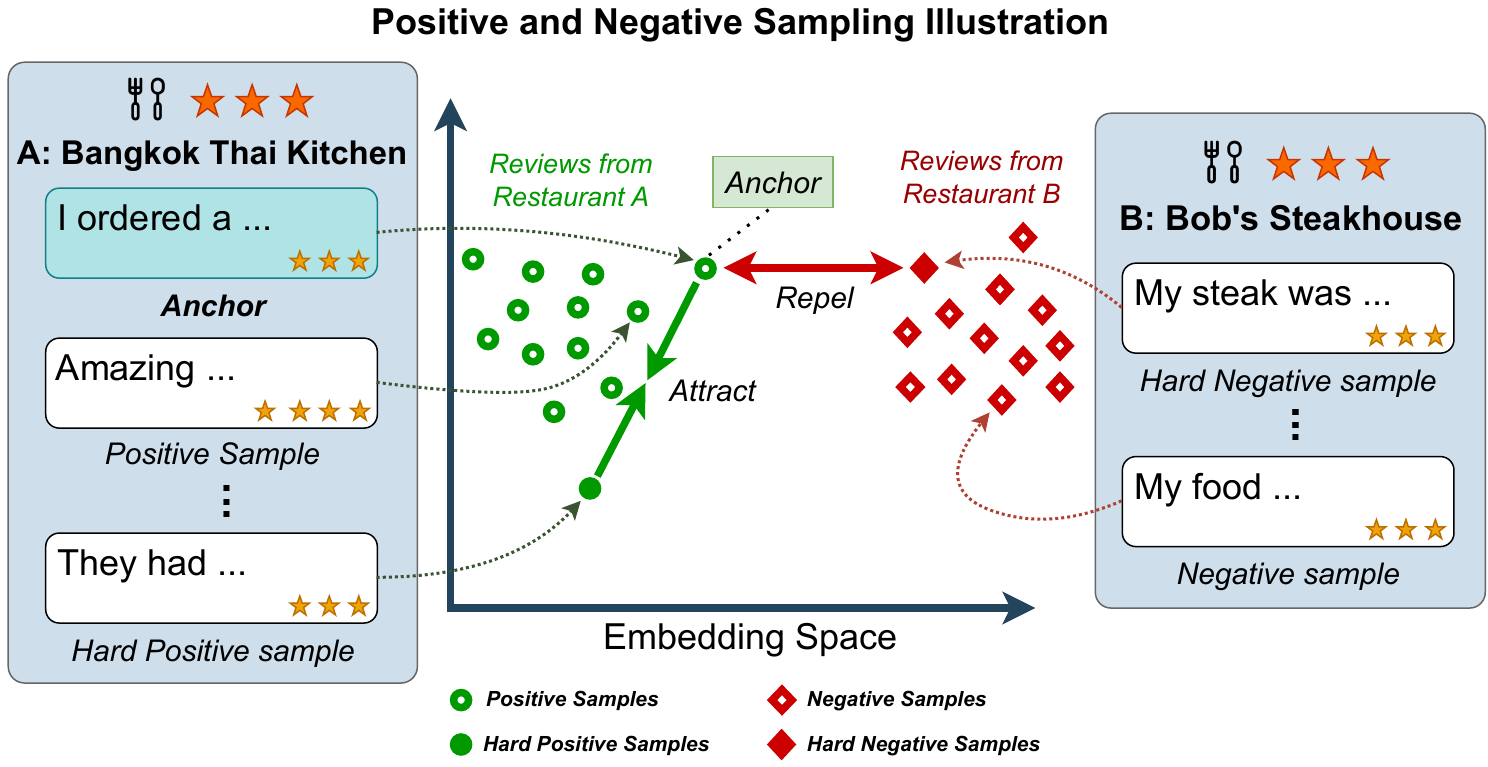}

\caption{
The objective of contrastive fine-tuning is to fine-tune the LM so that positive sample embeddings are closer to the anchor while negative sample embeddings are further from the anchor. Hard positive and hard negative samples are selected based on their distance to the anchor. A hard positive sample (e.g., the green filled circle) is a review from the same item which is furthest from the anchor.
A hard negative sample (e.g., the red filled diamond) is the closest review to the anchor but from a different item.
}
\vspace{-4mm}
\label{contrastive}
\end{figure*}
We now present our novel contrastive fine-tuning method for late fusion, in which review nuance is preserved during query scoring. In Contrastive Learning for Late Fusion RIR (CLFR), we fine-tune a language model $g$ (the embedding function) using the contrastive loss function in Equation (\ref{eq:totalLoss}) to produce embeddings of queries and reviews. The similarities of query and review embeddings are then evaluated using the dot product and aggregated into query-item scores by Equation (\ref{review_fusion}).

As opposed to single-level Neural IR contrastive sampling, RIR enables us to use the item-review structure in our sampling methods. Specifically, a positive pair $(x^A,x^+)$ is constructed from two reviews of the same item, while negative samples 
$\{x_i^-\}_{i=1}^{N-1}$
are obtained from reviews of different items. We explore several variations of sampling, including the use of item ratings, item keywords, and review embedding similarity, described below. While most of our methods use full reviews as samples, we also experiment with the use of review sub-spans for anchors; the goal of these variants is to reduce the discrepancy between anchor length and query length, since queries are typically much shorter than reviews. To keep training balanced, the same number of samples is used for all items.


\subsubsection{Positive Sampling methods} 
\label{positive_sampling}

$ \ $ 

    \subfour{I. Same Item (SI):} 
    We want two reviews from the same item to be close to each other in the embedding space. Therefore, if we have a review from item $i$ as an anchor $x^A$, the positive sample is a different review from item $i$, $x^+ \in \mathcal{D}_i$, sampled randomly unless otherwise mentioned.
    
    \subfour{II. Same Item, Same Rating (SI, SR):} 
   Building on SI, we further filter the positive pair reviews from the same item to have the same user rating as well. The motivation for this method is that reviews with the same rating are from people that had the same level of satisfaction from their experience. They are likely expressing a different aspect of their experience or a different phrasing of the same experience. This may be helpful for better embeddings of reviews according not only to language similarity but also user experience similarity.
    
    \subfour{III. Least Similar, Same Item (LS, SI):} Choosing a set of the hardest positive samples for the model to train on may help to further boost the performance. For the positive samples, we choose the samples that are from the same item and are furthest away from anchor in the BERT embedding space. Figure \ref{contrastive} (cf. green rings) shows this sampling method.
    
    \subfour{IV. Least Similar, Same Item, Same Rating (LS, SI, SR):} The reviews least similar to the anchor but from the same item and with the same rating are chosen as positive samples. This is a union of the previous methods.

  
\subsubsection{Negative Sampling methods} 
\label{negative_sampling}
$  \ $

\subfour{I. In-Batch (IB):} 
We use the In-Batch negative sampling method as explained in Section \ref{IB} for efficiency purposes. Since each anchor and positive sample pair in the mini-batch is from a different item, this ensures In-Batch negative samples for each anchor are from a different item than the anchor.


 \subfour{II. In-Batch + Hard Negatives (IB + HN):} In addition to the In-Batch negatives, we use the review that is most similar to the anchor $x^A$ --- but from a different item --- as a hard negative sample.
 We use dot product similarity to find the most similar review to each anchor in the (BERT) embedding space, as illustrated in Figure \ref{contrastive}, and hard negatives are cached from the fine-tuning procedure. By adding the hard negatives to the In-Batch negatives we aim to make the model more robust. 

\subsubsection{Data Augmentation}
\label{data_augmentation}
\textit{}

\noindent \\ We propose two data augmentation methods for RIR. Our first method exploits the meta-data tags that each item has, and our second method aims to mitigate the gap between self-supervised training and inference by shortening the anchor text.


\subfour{I. Prepending meta-data (PPMD):} Many item-review structures contain meta-data alongside textual and structural information. In our inference dataset, the items are restaurants that have categorical labels such as ``Pizza, Salad, Fast Food'' and usually include the cuisine type such as ``Mediterranean, Middle Eastern''. In order to use this meta-data without changing the model architecture, we prepend this data in text format to the review text during inference. This augmentation is done with the goal of better working with queries referring to categories and cuisine types.

\subfour{II. Anchor sub-sampling:} User queries are usually shorter than reviews. Thus, instead of taking a review as an anchor (which are the pseudo queries during the self-supervised fine-tuning), we take a shorter span from the review as the anchor in what we call sub-sampling anchor to a span (\textbf{SASP}). In an alternative method, we take a single sentence from the review as the anchor, and call this sub-sampling anchor to a sentence (\textbf{SASN}). For both cases, the span or sentence is chosen randomly, and we still use full reviews as positive and negative samples.

\subsection{CEFR: Contrastive Learning for Early Fusion RIR} \label{early} 
We now present our contrastive learning approaches for Early Fusion, in which item embeddings are learned before query-scoring. In Early Fusion, the query-item score $S_{q,i}$ is computed by Equation (\ref{ef_formula}) once we have an embedding $\vec{i}$ for each item $i$. A naive approach for obtaining the item embedding is to take the average of the review embeddings $\vec{r}_{i,k}$'s of item $i$. We call this naive approach Average EF.




We also introduce a new method which we call Contrastive Learning for Early Fusion RIR (CEFR), where we use contrastive learning to simultaneously fine-tune $g$ and learn an item embedding $\vec{i}$ in an end-to-end style. In particular, we let $\vec{i}$ be a learnable vector which we use instead of an anchor embedding $g(x^A)$ and initialize $\vec{i}$ with the naive item embedding obtained from Average EF. For a tuple $(\vec{i}, x^+, \{x^-_j\}_{j = 1}^{N-1})$, the loss given by Equation (\ref{npair}) thus becomes 
\begin{equation}
\label{ef_npair}
     L^{g}_{\mathrm{con}} (\vec{i}, x^+, \{x^-_j\}_{j = 1}^{N-1}) =  - \mathrm{log} \frac{e^{f(\vec{i},g(x^+))}}{e^{f(\vec{i},g(x^+))} + \sum_{j=1}^{N-1} e^{f(\vec{i},g(x^-_j))}}
\end{equation}
Given a set of training samples $\mathcal{S}$, with $s_i = (\vec{i},x^+,\{x^-_j\}_{j=1}^{N-1})$ representing the sample tuple for item $i$, and denoting $\vec{\mathcal{I}}$ as the set of learnable item embeddings, the objective of contrastive learning in CEFR is 
\begin{equation}
    \min_{\vec{\mathcal{I}},g} L^g(\mathcal{S}) = \min_{\vec{\mathcal{I}} ,g} \sum_{i = 1}^N L^{g}_{\mathrm{con}} (s_i)
\end{equation}


Note that CEFR simultaneously learns item embeddings and fine-tunes the language model $g$ (the embedding function), which is intended to allow flexibility in learning $\vec{i}$. For each item $i$, the positive sample $x_i^+ \in \mathcal{D}_i$ is a review from item $i$, and the negative samples are obtained via In-Batch negative sampling.

\section{Experiments}

\subsection{Reviewed-Item Retrieval Dataset (RIRD)}
\label{sec:dataset}

To address the lack of existing RIR datasets, we curated the \emph{Reviewed-Item Retrieval Dataset (RIRD)}\footnote{RIRD:  \url{https://github.com/D3Mlab/rir_data}} to support our analysis. We used reviews related to 50 popular restaurants in Toronto, Canada obtained from the Yelp dataset.\footnote{Yelp Dataset: \url{ https://www.yelp.com/dataset}}  
We selected restaurants with a minimum average rating of 3 and at least 400 reviews that were not franchises (e.g., McDonalds) since we presume franchises are well-known and do not need recommendations.
We created 20 queries for 5 different conceptual categories highlighted in Table \ref{tab:tablequeries} (with examples). These 5 groups capture various types of natural language preference statements that occur in this task.
We then had a group of annotators assess the binary relevance of each of these 100 queries to the 50 restaurants in our candidate set.
Each review was labeled by 5 annotators and the annotations showed a kappa agreement score of 0.528, demonstrating moderate agreement according to \citet{landis1977measurement}, which is expected given the subjectivity of the queries in this task \citep{balog2021interpretation}. There is a total number of 29k reviews in this dataset.

\begin{table}

\centering
  \caption{Categories of queries and their examples}
  \begin{tabular}{p{40mm}l}

    \hline
    \textbf{Query category} & \textbf{Example}\\

    \hline
    Indirect queries & \textit{I am on a budget}.\\
    Queries with negation & \textit{Not Sushi but Asian}.\\
    General queries & \textit{Nice place with nice drinks}.\\
    Detailed queries & \textit{A good cafe for a date that has live music}\\
    Contradictory queries & \textit{A fancy, but affordable place}.\\ 

  \hline
\end{tabular}

    \label{tab:tablequeries}
  
\end{table}

\subsection{Experimental Setup}
All experiments for fine-tuning models and obtaining semantic representations for reviews and 
queries were done with 8 Google TPUs. Each fine-tuning experiment took between 5-7 hours depending on the experimental setup. We use an Adam optimizer \cite{kingma2014adam} with a learning rate of $10^{-5}$ and use a validation set of 20\% of reviews for hyperparameter tuning. Each experiment was repeated 5 times with 5 different seeds to mitigate possible noise in dataset. We use a batch size of $N = 48$, and add one hard negative per tuple if hard negatives are used. The similarity function $f$ is the dot product. 
For the choice of LM, we use uncased base BERT \citep{bert-base-uncased}, and we select $K \in \{1,10,|\mathcal{D}_j|\}$ for Equation \ref{late_fusion_bg}, where $|\mathcal{D}_i|$ is the number of reviews for item $i$. Our code for reproducing all results is publicly available.\footnote{Code:  \url{https://github.com/D3Mlab/rir}} 
\subsection{Baselines}
\label{baselines_intro}

\subfour{Traditional IR:}
We use TF-IDF and BM25 with LF as  baselines. First, all query-review scores are computed by TF-IDF and BM25 ($b=0.75$, $k_1=1.6$) ranking functions. These scores are then fused in the same manner as in Section \ref{late_fusion_bg}, where query-review level information is aggregated to the query-item level using Equation (\ref{review_fusion}). Stopword removal and Lemmatization were also applied for these models using NLTK \citep{bird2009natural}.

\subfour{Masked Language Modeling (MLM):} Masked Language Modeling (MLM) is a primary self-supervised fine-tuning objective for many LMs such as BERT~\cite{devlin2018bert}. We use off-the-shelf BERT as a baseline and also fine-tune it with an MLM objective on our review data as another baseline. We train for 100 epochs with early stopping and a learning rate of $10^{-5}$.


\subfour{Neural IR models:} In order to compare state-of-the-art Neural IR models with our proposed models in the self-supervised regime, we use the following contrastively trained Neural IR language models: Contriever \citep{izacard2021towards}, Condenser \citep{gao2021condenser}  and CoCondenser \citep{gao2022unsupervised}.
We use publicly released pre-trained models for these baselines. Since only one model is available, confidence intervals are not applicable.

\subfour{IR-based contrastive:} We use contrastive learning with  self-supervised IR, explained in Section \ref{IRcon}, to fine-tune the LM as a baseline. The positive samples are created by ICT and IC from a single review, and In-Batch negative sampling is used. Since these methods are agnostic to the item associated with a review, as long as a review is not used for the positive pair, negative samples could come from reviews of the same item or reviews of different items. We use these baselines to examine the importance of using the two-level item-review structure in addition to the review text in our contrastive learning methods for RIR, since IR-based contrastive learning (ICT, IC) does not leverage that structure. 

\subsection{Evaluation Metrics}
We use Mean R-Precision (R-Prec) and Mean Average Precision (MAP) with 90\% confidence intervals to evaluate the performance of RIR in our experiments.
We note that R-Precision is regarded as a stable measure for averaging when the number of relevant items varies widely per query. 

\subsection{Results and Discussion}

This section studies four Research Questions (RQ) and discusses our experimental results for each. Specifically, we compare the effects of various sampling and data augmentation methods, evaluate Late Fusion against Early Fusion, and examine how our proposed methods compare to baselines.

\subfour{RQ1: Which sampling method from Section \ref{CLFR} gives the best performance for LM fine-tuning for LF?} The choice of sampling method for contrastive fine-tuning has an important effect on retrieval performance, but must be analysed jointly with the effect of $K$. In Table \ref{sampling}, we explore all the sampling options outlined in Section \ref{CLFR} with our base CLFR method using Same Item (SI) and In-Batch Negatives (IB) sampling. Regardless of $K$, there is no significant improvement from using Same Item (SI) Same Rating (SR) sampling over Same Item (SI) sampling. However, adding Least Similar (LS) to the base CLFR and having Same Rating (SR) provides an absolute and significant improvement of $0.028$ and $0.036$ for $K=1$ and $K=|\mathcal{D}_j|$ in R-Prec but no significant improvement for $K=10$. 
Adding hard negative samples fails to provide significant improvement regardless of the positive sampling method.

\subfour{RQ2: What is the impact of data augmentation methods from Section \ref{data_augmentation} for LM fine-tuning in LF setting?
} The results in Table \ref{sampling} show that prepending meta-data (i.e., cuisine type) to review text gives no significant improvement. We see that sub-sampling the anchors, by span or by sentence, achieves the best performance. On average, SASP and SASN  improve R-Prec on the base CLFR method by a significant amount 
for every $K$. This shows that being robust to the length of the anchor through sub-sampling plays a crucial role for contrastive learning in Neural RIR.  Due to space limitations, full comparative evaluation results are made available
in Appendix \ref{appendix}.

\subfour{RQ3: Does Early Fusion or Late Fusion work better for RIR?} Table \ref{tab:fusion} shows the performance of our Early and Late Fusion methods. We can see that our Late Fusion methods outperform Early Fusion significantly. We conjecture that since Early Fusion fuses all of an item's reviews into a single embedding vector, it may lose the nuance of individual review expressions or differences that averaged out in the Early Fusion process. In contrast, Late Fusion preserves the nuances of individual reviews until query-scoring.

\begin{table}[h]
        \caption{Results for exploring different techniques for CLFR with 90\% confidence intervals. Positive and negative sampling methods are explained in Section \ref{positive_sampling}. All rows have the base Same Item (SI) positive and In-Batch (IB) negative sampling. Additional sampling methods are specified in the first two columns.} 
    \resizebox{\columnwidth}{!}{%

    \begin{tabular}{@{}c|c|cc|cc|cc@{}}

    \hline
    \multicolumn{2}{c|}{\textbf{Sampling}} &  \multicolumn{2}{c|}{\textbf{$K$=1}}  &     \multicolumn{2}{c|}{\textbf{$K$=10}} &    \multicolumn{2}{c}{\textbf{$K$=$|\mathcal{D}_j|$ (Avg)}}  \\ \hline
    \textbf{Pos.}  &\textbf{Neg.}& \textbf{R-Prec}   & \textbf{MAP} & \textbf{R-Prec}   & \textbf{MAP} & \textbf{R-Prec}   & \textbf{MAP}   \\ \hline 
  -  & - & 0.497$\pm$0.010   &   0.569$\pm$0.021   &   0.514$\pm$0.011   &   0.587$\pm$0.013   &   0.495$\pm$0.017   &   0.575$\pm$0.018    \\ \hline 
  SR & - & 0.499$\pm$0.016   &   0.563$\pm$0.022   &   0.509$\pm$0.011   &   0.583$\pm$0.010   &   0.501$\pm$0.029   &   0.572$\pm$0.030    \\ \hline 
  LS & - & 0.513$\pm$0.014   &   0.589$\pm$0.018   &   0.500$\pm$0.014   &   0.584$\pm$0.015   &   0.487$\pm$0.008   &   0.569$\pm$0.013    \\ \hline 
  SR, LS & - &  0.525$\pm$0.004   &   0.596$\pm$0.014   &   0.515$\pm$0.013   &   0.590$\pm$0.016   &   \textbf{0.531$\pm$0.006}   &   \textbf{0.607$\pm$0.014}    \\ \hline 
  - & HN & 0.511$\pm$0.012   &   0.589$\pm$0.011   &   0.512$\pm$0.016   &   0.590$\pm$0.012   &   0.516$\pm$0.014   &   0.589$\pm$0.018    \\ \hline 
  SR, LS & HN & 0.504$\pm$0.015   &   0.571$\pm$0.006   &   0.516$\pm$0.016   &   0.594$\pm$0.012   &   0.517$\pm$0.012   &   0.596$\pm$0.013    \\ \hline 
  PPMD & - & 0.506$\pm$0.004   &   0.576$\pm$0.012   &   0.503$\pm$0.007   &   0.579$\pm$0.010   &   0.507$\pm$0.010   &   0.573$\pm$0.009    \\ \hline 
  SASP & - & 0.523$\pm$0.013   &   0.595$\pm$0.016   &   0.531$\pm$0.006   &   0.611$\pm$0.005   &   0.525$\pm$0.008   &   0.599$\pm$0.013    \\ \hline 
  SASN & - & \textbf{0.532$\pm$0.019}   &   \textbf{0.609$\pm$0.020}   &   \textbf{0.545$\pm$0.009}   &   \textbf{0.626$\pm$0.009}   &   \textbf{0.530$\pm$0.012}   &   \textbf{0.610$\pm$0.011}    \\ \hline 
    \end{tabular}
    }
 
 \label{sampling}  
\vspace{-6mm}
\end{table}
\begin{table}[h]
    \caption{Comparing our non-contrastive and contrastive Early Fusion (EF) methods with our base and best Late Fusion (LF) methods. Average EF takes an average BERT embedding of reviews as the item embedding. Base LF is CLFR with Same Item and In-Batch negatives. Best LF is CLFR with Same item and SASN and In-Batch negatives. We can see the noticeable improvement of using LF over EF. We conjecture this is due to the preservation of nuance in individual reviews during query scoring.}
    \centering
    \resizebox{\columnwidth}{!}{%
    \begin{tabular}{l|cc|cc|cc}\hline
        \multicolumn{1}{l|}{\textbf{EF Model}} &  \multicolumn{3}{c|}{\textbf{R-Prec}} & \multicolumn{3}{c}{\textbf{MAP}}    \\ \hline
        \multicolumn{1}{l|}{Average} & 
         \multicolumn{3}{c|}{0.297} & 
         \multicolumn{3}{c}{0.364}    \\ \hline
        \multicolumn{1}{l|}{CEFR} &  \multicolumn{3}{c|}{0.438$\pm$0.003} & \multicolumn{3}{c}{0.519$\pm$0.003}    \\ \hline \hline
        \multicolumn{1}{c|}{\textbf{}} &  \multicolumn{2}{c|}{\textbf{$K$=1}}  &     \multicolumn{2}{c|}{\textbf{$K$=10}} &    \multicolumn{2}{c}{\textbf{$K$=$|\mathcal{D}_j|$ (Avg)}}  \\ \hline
        \textbf{LF Model} & \textbf{R-Prec} & \textbf{MAP} & \textbf{R-Prec} & \textbf{MAP} & \textbf{R-Prec} & \textbf{MAP} \\ \hline
        Base LF    &  0.497$\pm$0.010   &   0.569$\pm$0.021   &   0.514$\pm$0.011   &   0.587$\pm$0.013   &   0.495$\pm$0.017   &   0.575$\pm$0.018    \\ \hline 
         Best LF  &  \textbf{0.532$\pm$0.019}   &   \textbf{0.609$\pm$0.020}   &   \textbf{0.545$\pm$0.009}   &   \textbf{0.626$\pm$0.009}   &   \textbf{0.530$\pm$0.012}   &   \textbf{0.610$\pm$0.011}    \\ \hline
    \end{tabular}
    }

    \label{tab:fusion}
\end{table}

\begin{table}[!]
\caption{CLFR (our best model) results on RIRD dataset versus baselines from section \ref{baselines_intro}. 
IR-based methods (Section \ref{IRcon}) and CLFR use In-Batch negatives (IB), and CLFR uses Same Item (SI) positive samples and SASN augmentation (Section \ref{data_augmentation}).}
\centering
\resizebox{\columnwidth}{!}{%
\begin{tabular}{@{}l|cc|cc|cc@{}} \cline{2-7} 
    & \multicolumn{2}{c|}{\textbf{$K$=1}}   & \multicolumn{2}{c|}{\textbf{$K$=10}} &    \multicolumn{2}{c}{\textbf{$K$=$|\mathcal{D}_j|$ (Avg)}} \\ \hline
    \textbf{Model}   &    \textbf{R-Prec} & \textbf{MAP} & \textbf{R-Prec} & \textbf{MAP} & \textbf{R-Prec} & \textbf{MAP} \\ \hline 
 
    TF-IDF    &   0.345  & 0.406 &   0.378  & 0.442 &   0.425  & 0.489 \\ \hline 
    BM25    &   0.393   & 0.450  &  0.417    & 0.490  &  0.421   & 0.495  \\ \hline 
    BERT    &   0.295  & 0.343 &   0.296  & 0.360 &   0.297  & 0.364 \\ \hline
    MLM    &   0.289  & 0.347 &   0.303  & 0.366 &   0.298  & 0.353 \\ \hline
    Condenser    &   0.358  & 0.410 &   0.390  & 0.449 &   0.378  & 0.428 \\ \hline 
    CoCondenser    &   0.445  & 0.505 &   0.481  & 0.553 &   0.482  & 0.570 \\ \hline 
    Contriever    &   0.375  & 0.427 &   0.418  & 0.482 &   0.458  & 0.519 \\ \hline 
    IR-based, IRC       &   0.355$\pm$0.024  & 0.422$\pm$0.033 &   0.355 $\pm$ 0.022  & 0.424$\pm$0.028 &   0.398$\pm$0.031  & 0.464$\pm$0.026 \\ \hline 
    IR-based, ICT        &   0.331$\pm$0.011  & 0.395$\pm$0.016 &   0.339$\pm$0.023  & 0.405$\pm$0.025 &   0.328$\pm$0.009  & 0.384$\pm$0.014 \\ \hline 
    \hline
    \textbf{CLFR}       &   \textbf{0.532$\pm$0.019}   &   \textbf{0.609$\pm$0.020}   &   \textbf{0.545$\pm$0.009}   &   \textbf{0.626$\pm$0.009}   &   \textbf{0.530$\pm$0.012}   &   \textbf{0.610$\pm$0.011}   \\ \hline 
\end{tabular} 
}

 \label{basline} 
\end{table} 
\vspace{-4mm}
\subfour{RQ4: How effective is our (best) method of using the structure of the review data for Late Fusion in RIR (CLFR) compared to existing baselines?} Table \ref{basline} compares the performance of state-of-the-art unsupervised IR methods with our best contrastive fine-tuning method for RIR. The out-of-domain Neural IR pre-trained language models CoCondenser and Contriever  perform noticeably better than base BERT, with CoCondenser performing the best among them. The traditional sparse retrieval models (TF-IDF and BM25) both outperform non-fine-tuned BERT, which emphasizes the importance of fine-tuning BERT for the downstream task of RIR. We also see that the self-supervised contrastive fine-tuning methods for IR (ICT and IRC) outperform non-fine-tuned BERT, but fall far behind the CoCondenser model. MLM fine-tuning of BERT also does not improve the performance of RIR, which is expected since it neither utilizes the structure of the data nor does this training objective directly support IR or RIR tasks. In contrast, by using the item-review data structure, our best method (CLFR with Same Item, SASN, and In-Batch Negatives) outperforms all contrastive IR, Neural IR, and sparse retrieval baselines.

\section{Conclusion}
In this paper, we proposed and explored novel self-supervised contrastive learning methods for both Late Fusion and Early Fusion methods 
that exploit the two-level item-review structure of RIR. 
We empirically observed that Same Item (SI) positive sampling and In-Batch negative sampling with sub-sampling anchor reviews to a sentence for Late Fusion achieved the best performance. This model significantly outperformed state-of-the-art Neural IR models, showing the importance of using the structure of item-review data for the RIR task. We also showed Late Fusion outperforms Early Fusion, which we hypothesize is due to the preservation of review nuance during query-scoring. Most importantly, this work opens new frontiers for the extension of self-supervised contrastive Neural-IR techniques to leverage multi-level textual structures for retrieval tasks. 


%
%
%
\bibliographystyle{splncs04nat}
\bibliography{refs.bib}

\appendix
\section{Appendix}
\label{appendix}
Table \ref{appendix_table} extends the experiments shown in table \ref{sampling} by combining more settings for contrastive learning. SASN augmentation still shows the best performance of $K=1,10$ and it performs best for $K$=$|\mathcal{D}_j|$ (Avg) while being combined with SR and LS. 
\begin{table}[h]

        \caption{Results for exploring different techniques for CLFR with 90\% confidence intervals. Positive and negative sampling methods are explained in Section \ref{positive_sampling}. All rows have the base Same Item (SI) positive and In-Batch (IB) negative sampling. Additional sampling methods are specified in the first two columns.} 
    \resizebox{\columnwidth}{!}{%

    \begin{tabular}{@{}c|c|cc|cc|cc@{}}

    \hline
    \multicolumn{2}{c|}{\textbf{Sampling}} &  \multicolumn{2}{c|}{\textbf{$K$=1}}  &     \multicolumn{2}{c|}{\textbf{$K$=10}} &    \multicolumn{2}{c}{\textbf{$K$=$|\mathcal{D}_j|$ (Avg)}}  \\ \hline
    \textbf{Pos.}  &\textbf{Neg.}& \textbf{R-Prec}   & \textbf{MAP} & \textbf{R-Prec}   & \textbf{MAP} & \textbf{R-Prec}   & \textbf{MAP}   \\ \hline 
 - & \  -  &  0.497$\pm$0.010   &   0.569$\pm$0.021   &   0.514$\pm$0.011   &   0.587$\pm$0.013   &   0.495$\pm$0.017   &   0.575$\pm$0.018    \\ \hline 
 SR   & \  -  &  0.499$\pm$0.016   &   0.563$\pm$0.022   &   0.509$\pm$0.011   &   0.583$\pm$0.010   &   0.501$\pm$0.029   &   0.572$\pm$0.030    \\ \hline 
 LS   & \  -  &  0.513$\pm$0.014   &   0.589$\pm$0.018   &   0.500$\pm$0.014   &   0.584$\pm$0.015   &   0.487$\pm$0.008   &   0.569$\pm$0.013    \\ \hline 
 SR, LS   & \  -  &  0.525$\pm$0.004   &   0.596$\pm$0.014   &   0.515$\pm$0.013   &   0.590$\pm$0.016   &   0.531$\pm$0.006   &   0.607$\pm$0.014    \\ \hline 
SI    & \   HN    &  0.511$\pm$0.012   &   0.589$\pm$0.011   &   0.512$\pm$0.016   &   0.590$\pm$0.012   &   0.516$\pm$0.014   &   0.589$\pm$0.018    \\ \hline 
 SR, LS   & \   HN    &  0.504$\pm$0.015   &   0.571$\pm$0.006   &   0.516$\pm$0.016   &   0.594$\pm$0.012   &   0.517$\pm$0.012   &   0.596$\pm$0.013    \\ \hline 
 PPMD   & \  -  &  0.506$\pm$0.004   &   0.576$\pm$0.012   &   0.503$\pm$0.007   &   0.579$\pm$0.010   &   0.507$\pm$0.010   &   0.573$\pm$0.009    \\ \hline 
SI, SASP    & \  -  &  0.523$\pm$0.013   &   0.595$\pm$0.016   &   0.531$\pm$0.006   &   0.611$\pm$0.005   &   0.525$\pm$0.008   &   0.599$\pm$0.013    \\ \hline 
 SASN   & \  -  &  \textbf{0.532$\pm$0.019}   &   \textbf{0.609$\pm$0.020}   &   \textbf{0.545$\pm$0.009}   &   \textbf{0.626$\pm$0.009}  &   0.530$\pm$0.012   &   0.610$\pm$0.011    \\ \hline 
 SR   & \ 
 HN   &  0.501$\pm$0.007   &   0.575$\pm$0.014   &   0.515$\pm$0.009   &   0.594$\pm$0.013   &   0.507$\pm$0.021   &   0.583$\pm$0.018    \\ \hline 
 LS   & \ 
 HN   &  0.513$\pm$0.016   &   0.588$\pm$0.014   &   0.509$\pm$0.021   &   0.586$\pm$0.019   &   0.500$\pm$0.029   &   0.576$\pm$0.024    \\ \hline 
 SR, LS, SASP   & \  -  &  0.520$\pm$0.016   &   0.599$\pm$0.012   &   0.523$\pm$0.013   &   0.608$\pm$0.015   &   0.521$\pm$0.010   &   0.590$\pm$0.005    \\ \hline 
 SR, LS, SASN   & \  -  &  0.528$\pm$0.013   &   0.603$\pm$0.012   &   0.536$\pm$0.012   &   0.614$\pm$0.015   &   \textbf{0.538$\pm$0.011}   &   \textbf{0.613$\pm$0.016}    \\ \hline 
 SR, LS, PPMD   & \  -  &  0.523$\pm$0.007   &   0.594$\pm$0.006   &   0.513$\pm$0.012   &   0.585$\pm$0.011   &   0.520$\pm$0.011   &   0.591$\pm$0.011    \\ \hline 
 SR, LS, SASP   & \ 
 HN   &  0.521$\pm$0.006   &   0.595$\pm$0.012   &   0.513$\pm$0.020   &   0.589$\pm$0.015   &   0.520$\pm$0.010   &   0.595$\pm$0.013    \\ \hline 
 SR, LS, SASN   & \ 
 HN   &  0.527$\pm$0.010   &   0.598$\pm$0.007   &   0.521$\pm$0.009   &   0.596$\pm$0.010   &   0.522$\pm$0.016   &   0.594$\pm$0.013    \\ \hline 
 SR, LS, PPMD   & \ 
 HN   &  0.513$\pm$0.012   &   0.584$\pm$0.009   &   0.514$\pm$0.011   &   0.586$\pm$0.003   &   0.513$\pm$0.015   &   0.584$\pm$0.009    \\ \hline 
 SASP   & \ 
 HN   &  0.514$\pm$0.011   &   0.591$\pm$0.010   &   0.521$\pm$0.018   &   0.594$\pm$0.017   &   0.528$\pm$0.013   &   0.604$\pm$0.008    \\ \hline 
 SASN   & \ 
 HN   &  0.506$\pm$0.015   &   0.589$\pm$0.014   &   0.520$\pm$0.014   &   0.600$\pm$0.016   &   0.519$\pm$0.017   &   0.598$\pm$0.019    \\ \hline 
 PPMD   & \ 
 HN   &  0.501$\pm$0.007   &   0.578$\pm$0.004   &   0.502$\pm$0.007   &   0.575$\pm$0.006   &   0.501$\pm$0.015   &   0.576$\pm$0.008    \\ \hline 
\end{tabular} 
}
 \label{appendix_table}  
\end{table}

\end{document}